\documentclass[pre,showpacs,superscriptaddress,showkeys]{revtex4}

\usepackage{graphicx}
\usepackage{dcolumn}
\usepackage{bm}
\usepackage{amssymb}

\begin{document}

\title{Relativistic Whistle: \\
High Order Harmonics Induced by \\
the Ultra-Intense Laser Pulse Propagating inside the Fiber}

\author{S. V. Bulanov}
\altaffiliation[Also at ]{Advanced Photon Research Center, JAERI, Kizu minami,
Kyoto-fu 619-0215, Japan} \email{bulanov@plasma.fpl.gpi.ru}
\affiliation{General Physics Institute RAS, Vavilov str. 38, Moscow 119991,
Russia}

\author{T. Zh. Esirkepov}
\email{timur@nonlin.mipt.ru}
\affiliation{Moscow Institute of Physics and Technology, Institutskij per. 9, Dolgoprudny Moscow region 141700, Russia}

\author{N. M. Naumova}
\affiliation{General Physics Institute RAS, Vavilov str. 38, Moscow 119991, Russia}

\author{I. V. Sokolov}
\affiliation{University of Michigan, Ann Arbor, MI 48109-2143, USA}

\date{May 22, 2002} 

\begin{abstract}
A propagation of an ultra-intense short laser pulse in a fiber
is investigated with two dimensional Particle-in-Cell simulations.
The fiber is a narrow hollow channel
with walls consisting of overdense plasma.
In the nonlinear interaction of the laser pulse with fiber walls
high order harmonics are generated.
Sufficiently high harmonics,
for which the fiber walls are transparent,
propagate outwards at certain angle.
This is a scheme of a
generator of ultra-short pulses of coherent light
with a very short wavelength.
\end{abstract}


\maketitle

\section{Introduction}

Generation of high order harmonics of the electromagnetic radiation during an
interaction of an ultra-intense laser pulse with underdense and overdense
plasmas is a manifestation of one of the most basic nonlinear processes in
physics. High order optical harmonics have been observed in the laser plasma
interaction for the laser intensity ranging from a moderate level to
relativistic intensities.

High order harmonics attract a great attention
due to a wide range of their applications for
the diagnostics, the UV and coherent X-ray sources,
lithography etc. (see Refs. \cite{NLO,H0,Review,Mourou}).
Recently,
the high order harmonics polarization properties have been used in
Ref. \cite{Norreys} to detect the strongest magnetic field
generated in the laser plasmas.

Different physical mechanisms of high order harmonics generation
have much in common because of
a property of nonlinear systems to react in an anharmonical manner
to a periodic driving force.
On the other hand, specific realization of this property depends on the
circumstances of the laser-matter interaction and
mainly on the laser intensity.
At moderate intensities, generation of high order harmonics
is caused by an atom anharmonic response to an oscillating electric field.
At relativistic intensities,
the electron quiver energy becomes higher than the rest mass energy
and the Relativistic Nonlinear Optics comes into play \cite{Mourou}.
In this regime high order harmonics generation
is due to a nonlinear dependence of the particle mass on its momentum and
due to electron density modulations in the electromagnetic wave.
In the present paper we discuss high harmonic generation
in collisionless plasmas by a relativistically strong electromagnetic wave.

In the underdense plasmas an ultra-intense laser pulse causes high harmonics
generation through a parametric excitation of electromagnetic and electrostatic
waves with different frequencies. In an interaction with overdense plasmas the
laser radiation (partially) reflects back at the plasma-vacuum interface, in
the case of sharp plasma boundary, or at a critical density surface, in the
case of gradual density profile. The reflecting layer in the plasma is dragged
by the electromagnetic wave back and forth in the direction of the incident
light, as well as in the transverse direction,
forming an oscillating mirror (see Refs. \cite{BNP,PG,LMP,VNPB,M1,Zepf-98,Tarasevich}).
The light reflected by the oscillating mirror
contains odd and even harmonics,
whose polarization and amplitudes depend on the incidence angle,
intensity and a polarization of the laser pulse.

High efficiency of the laser energy transformation into the energy of high
harmonics can be achieved in a scheme, in which the laser pulse propagates
inside a hollow channel \cite{BKPP}. In this case, if the channel is
sufficiently long, the laser pulse undergoes a multiple reflection at the
plasma-vacuum interface, and a portion of the laser pulse energy transforms
into the energy of high harmonics at each reflection event. If the laser pulse
propagates inside a hollow channel in a tubiform fiber, the radiation with
sufficiently high frequency propagates through the fiber walls. The fiber walls
are transparent for harmonics, whose indices are greater than some critical
value. Only these harmonics are emitted from the fiber at certain angle with
respect to the fiber axis. This approach allows one to design a high frequency
radiation source with controlled properties, which depend on the fiber diameter
and the fiber wall thickness. We call this source ``the relativistic whistle''.

In this paper we investigate a propagation of
an ultra-intense short laser pulse in a fiber,
with an aim to study the properties of the high harmonic radiation.
Here we use the two dimensional approximation,
leaving the investigation of the 3D problem
for forthcoming publications.

\section{Matching conditions}

In the two-dimensional case the fiber corresponds to
two thin foils of a finite length.
The foils are parallel to each other, and to the x-axis in the $x,y$-plane.
Let the wall thickness of the foil is $l$,
and a distance between the foils equal to $L$.
We consider the foils to be made of the collisionless overdense
plasma with the electron density $n_e$.
Such a configuration can be formed
by an ultra-intense femtosecond laser pulse
interacting with the solid density fiber,
when the fiber wall material is ionized during the first
half-period of the laser pulse,
and the length of the laser pulse is much
shorter than the timescale of a hydrodynamic expansion of the plasma.
In the framework of this approximation we can assume the ions to be at rest.

From the dispersion equation, $\omega^2=k^2c^2$, where $\omega$ is the wave
frequency and $k^2=k^2_x+k^2_y$, we find the group velocity of the wave inside
the channel, $v_{gr}=\partial \omega/\partial k_x =
(c/\omega)(\omega^2-\pi^2/L^2)^{1/2} = c \cos \theta$, where $\tan
\theta=k_y/k_x=(4 L^2/\lambda^2-1)^{-1/2}$ and $\lambda=2\pi c/\omega$. The
matching conditions for the wave inside the channel, and the wave radiated
outwards, show that all the outgoing waves propagate at the angle $\theta$ with
respect to the channel axis, independently of the wave frequency.

The fact that all the outgoing waves have the same angle of propagation can be
demonstrated in a different way. After the Lorentz transform into the reference
frame moving at a speed of $v_{gr}$ along the $x-$axis, the $x-$component of
the fundamental mode wave vector vanishes: $k_{x}^{\prime }=0$ (see for details
Refs. \cite{BNP,VNPB,Bourdier}). The problem becomes
one-dimensional. Therefore high frequency waves excited by the fundamental mode
at the plasma-vacuum interface also have zero $x-$component of the wave vector
in this reference frame. They propagate outwards in the direction perpendicular
to the fiber walls. Performing the inverse transform back to the laboratory
reference frame, we obtain that all the outgoing waves propagate at the same
angle $\theta $ with respect to the channel axis, independently of the wave
frequency. Their energy is localized inside a slab co-moving with the laser
pulse along the $x-$axis. The slab length is equal to the length of the laser
pulse. Its transverse size depends on the effective length of the interaction.
According to the selection rules of the harmonic generation at each wall (see
Refs. \cite{BNP,LMP,VNPB}), the $s-$polarized fundamental mode in
the fiber generates $s-$polarized odd harmonics and $p-$polarized even
harmonics of radiation. The $p-$polarized fundamental mode generates only
$p-$polarized harmonics of radiation.

The amplitude of the $n$-th harmonic can be found in
the approximation of an infinitely thin walls of the fibre (see Ref. \cite{VNPB}).
It is $E^{(n)}=K^{(0,n)}E^{(0)}$,
where $E^{(0)}$ and $E^{(n)}$ are amplitudes
of the fundamental mode and of the $n$-th harmonic, respectively.
The transformation coefficient $K^{(0,n)}$
is given by
$K^{(0,n)}\approx \left[(a_0/\epsilon) \sin \theta\right]^n$,
in the limit $n \omega/\omega_{pe}\ll\varepsilon$.
Here $a_{0}=eE^{(0)}/m_{e}\omega _{0}c$ is
the dimensionless amplitude of the fundamental mode,
$\varepsilon =2\pi n_{0}e^{2}l/m_{e}\omega _{0}c$
is the dimensionless parameter (relativistic transparency measure),
$\omega _{0}$ is the frequency of the fundamental mode.

\section{Parameters of the Computer Simulations}

\begin{figure}
\includegraphics{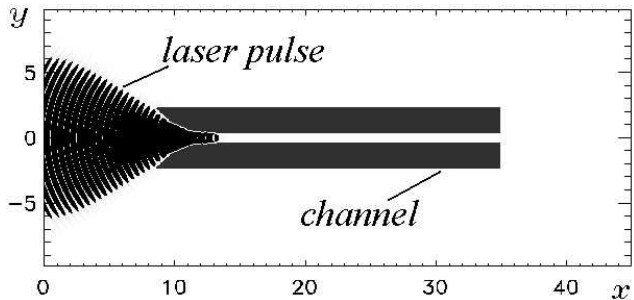}
\caption{\label{fig1}
The electromagnetic energy density and the plasma
density distribution at $t=15 \times 2 \pi/\omega_0$ when the pulse enters
into the fiber.}
\end{figure}

We perform computer simulations using the two-dimensional version of
Particle-in-Cell relativistic electromagnetic code REMP, based on `density
decomposition' scheme \cite{Esirkepov}. Computation box has a size $45\lambda
\times 20\lambda $. Since the problem of high harmonic generation requires
substantially high spatial resolution, the computation mesh has 50 cells per
$\lambda $. The number of quasiparticles per cell is equal to 16 in the plasma
region. The channel walls have a length $l_{ch}=25\lambda $ and width $2\lambda
$, a distance between the walls is $L=0.7\lambda $. The ions are assumed to be
immobile and the electron density is equal to $8n_{cr} $, where the critical
density is $n_{cr}=\omega _{0}^{2}m_{e}/4\pi e^{2}$. Since the channel diameter
is less than the laser light wavelength, only the $s$-polarized electromagnetic
wave (whose electric field is directed along the $z$-axis) can propagate inside
the channel. For the parameters chosen, only harmonic, whose index is higher
than 4, can propagate through the walls. This conclusion follows from the
boundary conditions for the electromagnetic wave, which is obliquely incident
on the plasma-vacuum interface, at the angle $\theta$. The $s$-polarized laser
pulse is initialized in a vacuum region at the left hand side from the channel
entry. The pulse aperture is f/1, its length is equal to $7.5\lambda $, and its
dimensionless amplitude at the $1\lambda$ focus spot is $a_{0}=3$, which
corresponds to the intensity $I\approx 10^{19}$W/cm$^{2}$. These parameters are
close to the parameters of the laser in the CUOS at the University of Michigan
\cite{Mourou}. We notice here that the pulse focusing into the $1\lambda$ spot
is discussed in Ref. \cite{Mourou}. In our simulation model the entry of the
channel is smoothed to decrease the laser pulse reflection. Fig. \ref{fig1} shows the
focusing laser pulse and the fiber. We see that the smoothing of the channel
entry provides almost reflectionless matching of the laser pulse and the
channel.

\section{Results of PIC simulations}

\begin{figure}
\includegraphics{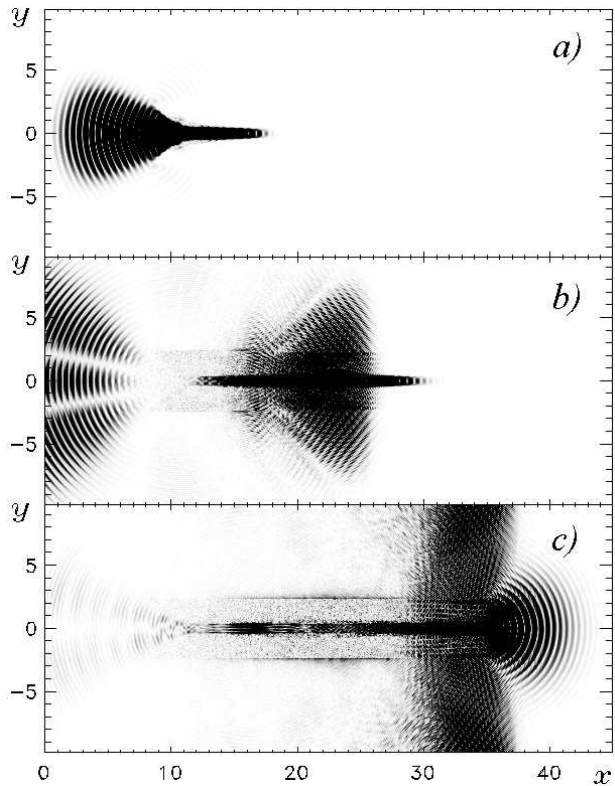}
\caption{\label{fig2}
The electromagnetic energy density distribution in the $x,y$ plane
at $t=20\times 2\pi /\omega $ in frame (a),
   $t=35\times 2\pi /\omega $ in frame (b), and
at $t=50\times 2\pi /\omega $ in frame (c).
}
\end{figure}

When the laser pulse propagates inside the channel a portion of its energy
is reflected back due to the process similar to the stimulated Raman
scattering in the plasma of the channel walls; a portion of its energy is
transformed into the energy of high harmonic waves radiated outwards; the
remaining part of the laser pulse propagates through the channel toward the
end of the fiber, then it goes out as a diverging electromagnetic wave.
These successive phases are seen in Fig. \ref{fig2}, where we present the
electromagnetic energy density distribution in the $x,y-$plane
at different times.

\begin{figure}
\includegraphics{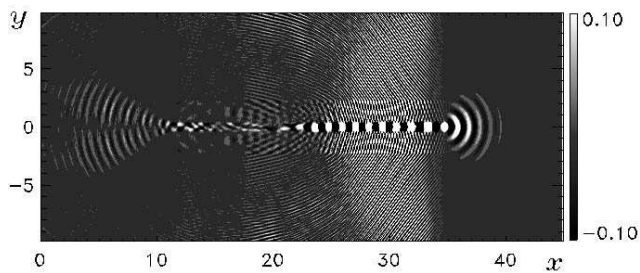}
\caption{\label{fig3}
The distribution of the $z$-component of the electric field
in the $x,y$-plane at $t=45 \times 2 \pi/\omega_0$.
}
\end{figure}

In Fig. \ref{fig3} we show the distribution of the $z$-component of the electric
field in the $x,y$-plane.
We see the laser pulse inside the channel and
the high harmonics aside the channel.
In this figure all the harmonics have odd indices
and they are $s$-polarized, as noticed above.
The spatial region filled by high harmonics
has a length equal to the laser pulse length and
the transverse size (along the $y$-axis) equal to
$\delta tc\sin \theta $, where
$\sin \theta \approx 1/2$ and
$\delta t\approx 20\times 2\pi /\omega _{0}$.

\begin{figure}
\includegraphics{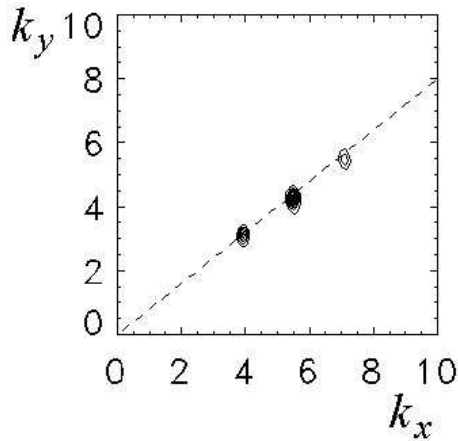}
\caption{\label{fig4}
Spectrum of the $s$-polarized harmonics.
The dashed line denotes the dependency $k_y\approx 0.8 k_x$.
Each peak corresponds to a harmonic with index $n=(k_x^2+k_y^2)^{1/2}$.
}
\end{figure}

We perform the spatial Fourier transform of the $z$-component of the electric
field inside a sub-domain ($18\lambda <x< 28\lambda $; $3\lambda <y< 10\lambda
$) at $t=30 \times 2 \pi/\omega_0$. The resulting spectrum of the $s$-polarized
light inside this sub-domain is presented in Fig. \ref{fig4}. We see the odd harmonics
with the harmonic index $n=(k_x^2+k_y^2)^{1/2}\geq 5$. The spectrum peaks are
arranged along the straight line given by $k_y\approx 0.8 k_x$, i. e. the angle
of propagation is smaller than $\arccos(1-\pi^2 c^2/\omega_0^2 L^2)^{1/2}$. It
means that the effective channel width is larger than $0.7\lambda $. The
increase of the channel width is apparently due to the action of the
ponderomotive force on the electrons at the plasma-vacuum interface.

From the expression $m_{e}c^{2}\nabla (1+a^{2})^{1/2}\approx 4\pi
n_{0}e^{2}\delta y$, which corresponds to the balance between the ponderomotive
force and the force due to electric charge separation, we find that the channel
width becomes larger by $2\delta y\approx 0.1\lambda $. Assuming now the
channel width to be equal to $0.8\lambda $ we obtain the propagation angle
$\theta _{eff}\approx \arctan(0.8)\approx 38.7^{o}$, in agreement with the
dependence seen in Fig. \ref{fig4}.

\begin{figure}
\includegraphics{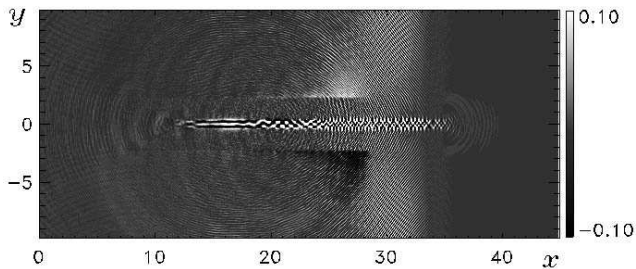}
\caption{\label{fig5}
The distribution of the $z$-component of the magnetic field
in the $x,y$-plane at $t=45 \times 2 \pi/\omega_0$.
}
\end{figure}

In Fig. \ref{fig5} we present the distribution of the $z$-component of the magnetic
field at the same point in time as Fig. \ref{fig3}.
We see the high harmonics with even indices.
They are $p$-polarized.
As noticed above, the fundamental mode of the laser pulse inside the channel
does not contain the $p$-polarized component and we see only the high
harmonics inside and aside the channel, as well as the low frequency surface
mode in the vicinity of the outer plasma-vacuum interfaces on both sides of
the fiber. The even-index high harmonics outside the channel are
localized in the same region as the odd-index harmonics.

\begin{figure}
\includegraphics{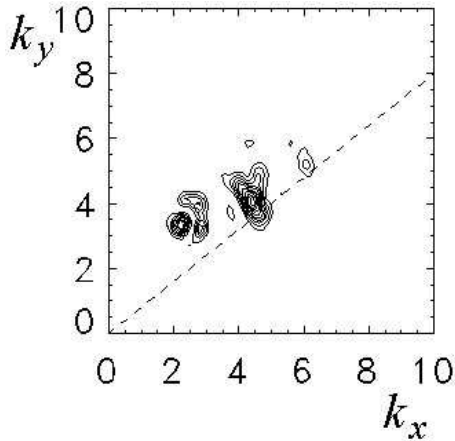}
\caption{\label{fig6}
Spectrum of the $p$-polarized harmonics.
}
\end{figure}

The spectrum of the even-index high harmonics calculated for the $z$-component
of the magnetic field in the same sub-domain and at the same point in time as
in the case of the odd-index harmonics is presented in Fig. \ref{fig6}. We see the even
harmonics with the harmonic index $n\geq 6$. As in the previous figure, the
spectrum peaks are arranged along the same line $k_y\approx 0.8 k_x$.

\section{Conclusion}

The high harmonic generation in the interaction of the ultra-intense short
laser pulse with a narrow fiber is demonstrated with PIC simulation. Due to
nonlinear interaction of the laser pulse with the fiber walls, the quivering
back and forth electrons form the oscillating mirrors. The high order harmonics
are generated as a result of a multiple reflection of the laser light from the
oscillating mirrors. The fiber walls are opaque for the electromagnetic wave
with a frequency below some critical value, which is defined by the electron density of
the fiber walls and the channel width. As a result, only sufficiently high
harmonics, for which the fiber walls are transparent, propagate outwards at
certain angle, independently of the harmonic index. In addition, we observe the
excitation of the $p$-polarized low frequency mode, which is localized at the
outer surface of the fiber. The longitudinal scale of the low frequency mode is
of the order of the laser pulse length. It provides an example of the nonlinear
rectification of the light. The high harmonics energy is localized in the
region with a length (along the $x$-axis) equal to the laser pulse length,
and the width (along the $y$-axis) equal to $l_{ch}\tan \theta _{eff}$. This region
is co-moving with the fundamental mode inside the fibre. The amplitude of the
$n$-th harmonic can be estimated as $(a_{0}\sin \theta _{eff})^{n}(\omega
_{0}/\omega _{pe})^{1+2n}$, where $\omega _{pe}=(4\pi n_{0}e^{2}/m_{e})^{1/2}$.
For $a_{0}=3$, $(\omega_{0}/\omega _{pe})^{2}=1/8$ and $\theta _{eff}\approx
38^{o}$, we obtain $E^{(n)}/E^{(0)}\approx 0.35(0.23)^{n}$. This proves an
effective mechanism for generation of ultra-short pulses of coherent light with
a very short wavelength. The corresponding device can be called ``the
relativistic whistle''.

\end{document}